\def\year{2020}\relax
\documentclass[letterpaper]{article} 
\usepackage{aaai20}  
\usepackage{times}  
\usepackage{helvet} 
\usepackage{courier}  
\usepackage[hyphens]{url}  
\usepackage{graphicx} 
\usepackage{graphicx}  
\usepackage{booktabs}
\title{Image Formation Model Guided Deep Image Super-Resolution}
\author{
Jinshan Pan,\textsuperscript{\rm 1}
Yang Liu,\textsuperscript{\rm 2}
Deqing Sun,\textsuperscript{\rm 3}
Jimmy Ren,\textsuperscript{\rm 4}\\
{\bf \Large
Ming-Ming Cheng,\textsuperscript{\rm 5}
Jian Yang,\textsuperscript{\rm 1}
Jinhui Tang\textsuperscript{\rm 1}}\\
\textsuperscript{\rm 1}Nanjing University of Science and Technology,
\textsuperscript{\rm 2}Dalian University of Technology,\\
\textsuperscript{\rm 3}Google,
\textsuperscript{\rm 4}SenseTime Research,
\textsuperscript{\rm 5}Nankai University\\
\url{https://github.com/jspan/PHYSICS_SR}
}
 
\begin{document}

\maketitle

\begin{abstract}
   We present a simple and effective image super-resolution algorithm that imposes an image formation constraint on the deep neural networks via pixel substitution.
   The proposed algorithm first uses a deep neural network to estimate intermediate high-resolution images, blurs the intermediate images using known blur kernels,
   and then substitutes values of the pixels at the un-decimated positions with those of the corresponding pixels from the low-resolution images.
   The output of the pixel substitution process strictly satisfies the image formation model and is further refined by the same deep neural network in a cascaded manner.
   The proposed framework is trained in an end-to-end fashion and can work with existing feed-forward deep neural networks for super-resolution and converges fast in practice.
   Extensive experimental results show that the proposed algorithm performs favorably against state-of-the-art methods. The training code and models are available at \url{https://github.com/jspan/PHYSICS_SR}.
\end{abstract}

\section{Introduction}
Single image super-resolution (SR) aims to estimate a high resolution (HR) image from a low resolution (LR) image.
It is a classical image processing problem and has received active research efforts in the vision and graphics community within the last decade.
The renewed interest is due to the widely-used high-definition devices in our daily life, such as iPhoneXS ($2436\!\times\!1125$), Pixel3 ($2960\!\times\!1440$), iPad Pro ($2732\!\times\!2048$), SAMSUNG Galaxy note9 ($2960 \times 1440$), and 4K UHDTV ($4096 \times 2160$).
There is a great need to super-resolve existing LR images so that they can be pleasantly viewed on high-definition devices.

\begin{figure}[t]
\centering
\begin{tabular}{c}
\includegraphics[width=0.98\columnwidth]{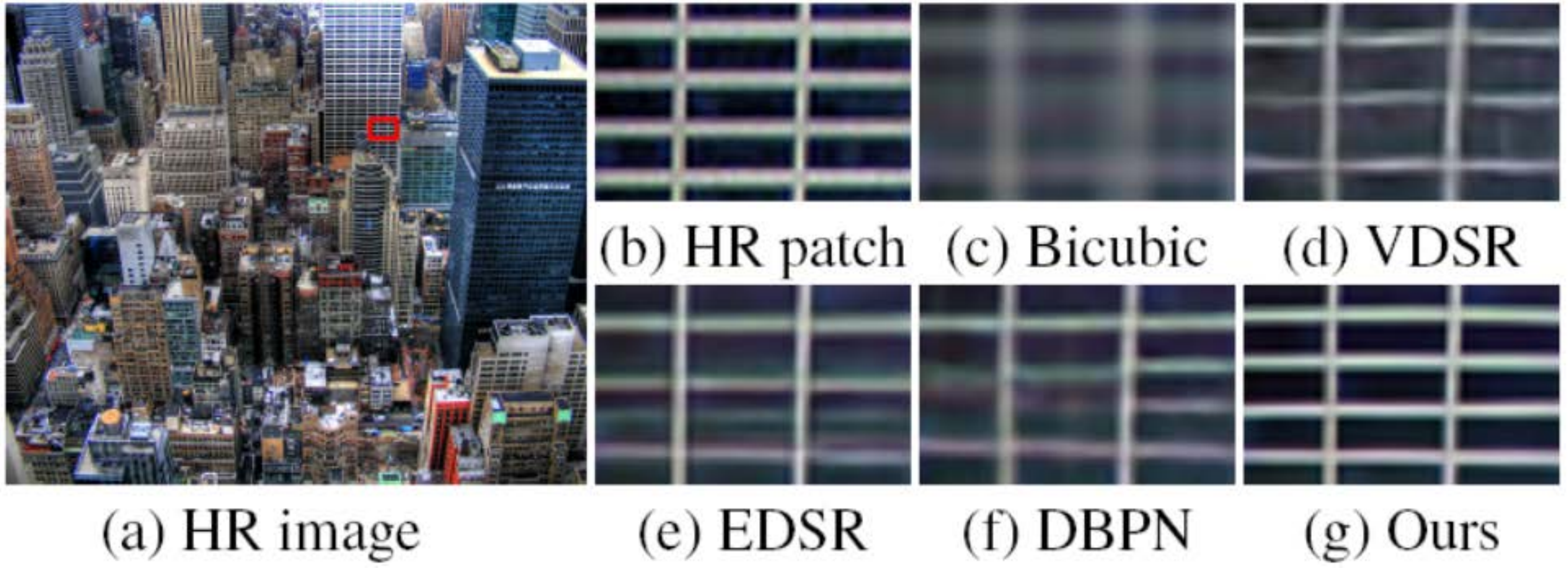}
\end{tabular}
\caption{Super-resolution result ($\times 4$). Our algorithm uses the image formation of super-resolution to constrain a deep neural network via pixel substitution, which generates the images satisfying the image formation model and better recovers structural details.
}%
\label{fig: teaser}
\end{figure}

Recently significant progress has been made by using convolutional neural networks (CNNs) in a regression way.
For example, numerous methods~\cite{SRCNN/ECCV,VDSR,DRCN,ESPCN,edsr,SRGAN,Accelerating/dong,Zhang_2018_CVPR} develop feed-forward networks with advanced network architectures (e.g., residual network~\cite{ResNet/CVPR16}, attention model~\cite{CAN/eccv}) or optimization strategies to learn the LR-to-HR mapping.
These methods are efficient and outperform conventional hand-crafted prior-based methods by large margins.
However, as the SR problem is highly ill-posed, using feed-forward networks may not be sufficient to estimate the LR-to-HR mapping.
In particular, the reconstructed HR images often do not strictly satisfy the image formation model of SR.

To address this issue, several methods improve feed-forward networks with feedback schemes,
such as re-implementing iterative back-projection method~\cite{IBP} by deep CNNs~\cite{DBPN}, using deep CNNs as image priors to constrain the solution space in a variational setting~\cite{prior/cnn/sr/hq},
using the image formation model in a feedback step to constrain the training process~\cite{physicsgan}.
However, these algorithms all regenerate LR images from the reconstructed intermediate HR results.
The downsampling operation leads to information loss and thus makes these algorithms hard to estimate the details and structures (e.g., Figure~\ref{fig: teaser}(f)).

We note that the LR image is usually assumed to be obtained by a convolution followed by a downsampling process on the HR image.
Under this assumption, at the un-decimated positions, the LR image should have the same pixel values as the blurred HR image which is obtained by applying a convolution operation to the clear HR image.
Thus, we should impose this image formation constraint in the network architecture to generate high-quality images.

However, it is challenging to apply the hard image formation constraint to deep neural networks, because it requires a feedback loop.
To this end, we propose a cascaded architecture to efficiently learn the network parameters.
The algorithm first generates an intermediate HR image by a deep neural network and then uses the LR image
to update the intermediate HR image based on the image formation process.
The updated intermediate HR image is further refined by the same deep neural network.
Extensive experiments show that the proposed algorithm based on this cascaded manner converges quickly and can generate high-quality images with clear structures.

\section{Related Work}

We briefly discuss methods most relevant to this work and refer interested readers to \cite{chihyuan/eccv/YangMY14} for comprehensive reviews.

The method by \cite{SRCNN/ECCV} is the first to develop a CNN method for SR, named as SRCNN.
Kim et al. show that the SRCNN algorithm is less effective at recovering image details and propose a residual learning algorithm using a 20-layer CNN~\cite{VDSR}.
In~\cite{DRCN}, Kim et al. introduce a deep recursive convolutional network (DRCN) using recursive-supervision and skip connections.
The recursive learning algorithm is further improved by~\cite{DRRN}, where both global and local learning are used to increase the performance.
However, these methods usually upscale LR images to the desired spatial resolution using bicubic interpolation as input to a network, which is less effective for the details restoration
as the bicubic interpolation method usually removes details~\cite{ESPCN}.

As a remedy, the sub-pixel convolutional layer~\cite{ESPCN} or deconvolution layer~\cite{Accelerating/dong} are developed based on SRCNN.
In~\cite{LapSR}, the Laplacian Pyramid Super-Resolution Network (LapSRN) is proposed to predict sub-band residuals on various scales progressively.
Based on the sub-pixel convolutional layer, several algorithms develop the networks with advanced architectures and strategies, e.g., dense skip connection~\cite{Dense-net/tong,Zhang_2018_CVPR}, dual-state recurrent models~\cite{Dual/State}, residual channel attention method~\cite{CAN/eccv}.
These algorithms are effective for super-resolving LR images but usually tend to smooth some structural details.
To generate more realistic images, Generative Adversarial Networks (GANs) with both pixel-wise and perceptual loss functions have been used to solve the SR problem~\cite{SRGAN,Sajjadi_2017_ICCV}.
Recent work~\cite{GAN/data/eccv18} first uses GANs to generate more realistic training images then trains GANs with the generated training images for SR.
Motivated by the generative network in~\cite{SRGAN}, the method by~\cite{edsr} removes some unnecessary non-linear active functions in the generator~\cite{SRGAN} and propose an Enhanced Deep Super-Resolution (EDSR) network to super-resolve images.
However, all these methods directly predict the nonlinear LR-to-HR mapping based on feed-forward networks.
They do not explore the domain knowledge of the SR problem and tend to fail at recovering fine image details.

%
To generate high-quality images that satisfy the image formation constraint, Wang et al. propose a sparse coding network (SCN) based on the sparse representation prior~\cite{Wang/2015/ICCV}.
In~\cite{prior/cnn/sr/hq}, Zhang et al. learn a CNN as an image prior to constrain the iterative back-projection algorithm~\cite{IBP}.
More recently, the deep neural networks with feedback schemes have been used in SR. Haris et al. improve the conventional iterative back-projection algorithm using CNNs~\cite{DBPN}.
Pan et al. propose a GAN model with an image formation constraint for image restoration~\cite{physicsgan}.
However, these algorithms need to regenerate low-resolution images in the feedback step which accordingly increase the difficulty for the details and structures restoration.
Moreover, the image formation in these methods is used as a soft constraint, which does not directly help the SR results~\cite{physicsgan}.
Using the image formation as a hard constraint is first introduced by~\cite{shan/sr/sa08} in the variational framework.
This method~\cite{shan/et/al} uses the pixel substitution to ensure that the generated SR results satisfy the image formation of SR in a hard way.
However, it cannot effectively recover the details and structures as only the sparsity of gradient prior is used.

In this work, we revisit the idea of pixel substitution to impose the hard image formation constraint in a deep neural network.
The proposed algorithm explores the information from both HR images and LR inputs by a deep neural network in a regression way
and is able to generate the results satisfying the image formation model, thus facilitating the high-quality image restoration.
%
\begin{figure*}[!t]
\centering
\begin{tabular}{c}
\includegraphics[width = 0.95\linewidth]{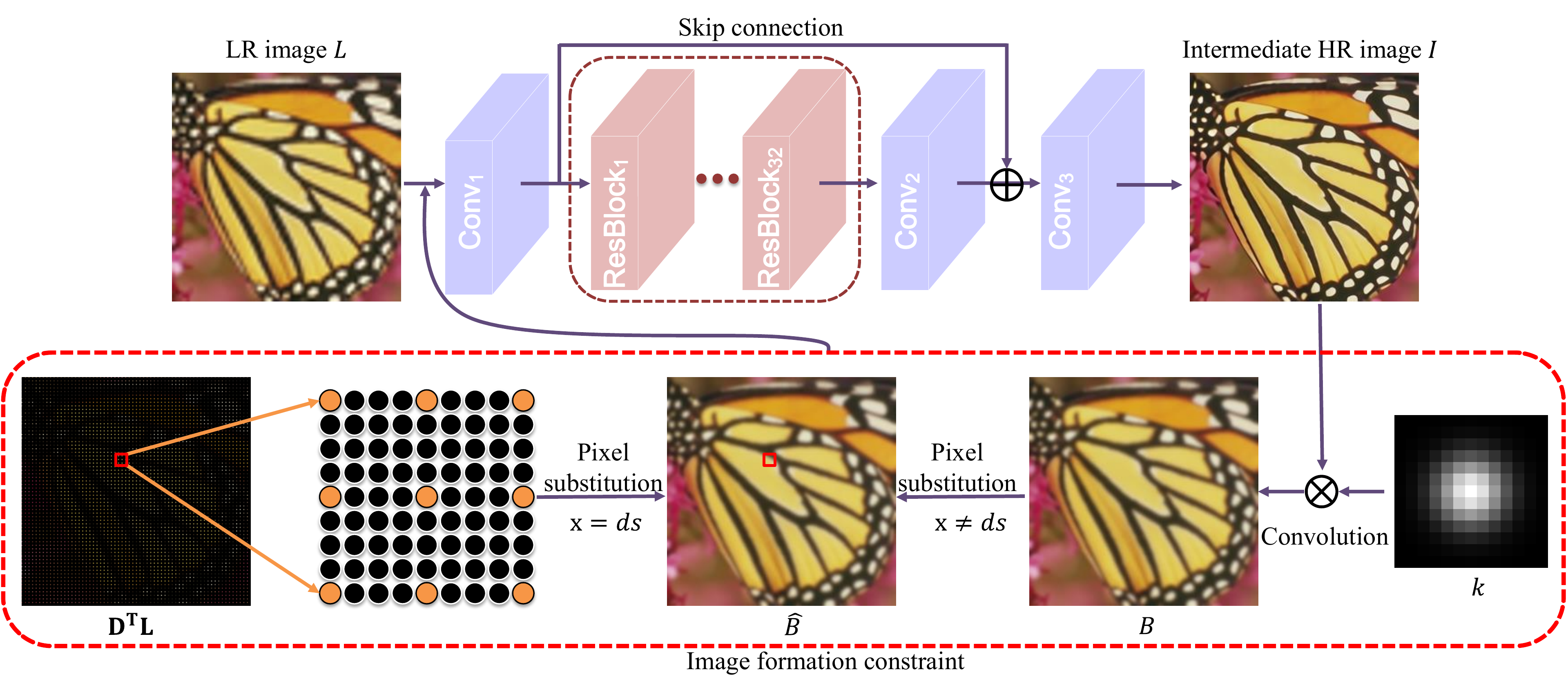}\\
\end{tabular}
\caption{An overview of the proposed method. The image formation constraint is enclosed in the dotted red box, which is used to constrain a deep CNN for super-resolution.
At each stage, our algorithm first generates an intermediate HR image ${I}$ by a deep CNN model and updates the intermediate HR image ${I}$ according to the image formation of SR by the pixel substitution~(\ref{eq:pxiel:substitution}). The updated image $\hat{B}$ is then taken as an input for the next stage.
The network is solved in a cascaded manner and generates better high-quality images.
}
\label{fig: flow-chart}
\end{figure*}

\section{Image Formation Process}
We first describe the image formation process of the SR problem and then derive the image formation constraint.
Given a HR image $I$, the process of generating the LR image $L$ is usually defined as
\begin{equation}
L =\downarrow^{s}(k\otimes I),
\label{eq: sr-formation-convolution}
\end{equation}
where $k$ denotes the blur kernel, $\otimes$ denotes the convolution operator, and $\downarrow^{s}$ denotes the downsampling operation with a scale factor $s$.
Mathematically, this image formation process can be rewritten as
\begin{equation}
\mathbf{L} =\mathbf{DK}\mathbf{I},
\label{eq: sr-formation-2}
\end{equation}
where $\mathbf{K}$ denotes the filtering matrix corresponding to the blur kernel $k$; $\mathbf{D}$ denotes the downsampling operation; $\mathbf{L}$ and $\mathbf{I}$ denote the vector forms of $L$ and $I$.

Applying the upsampling matrix, i.e., $\mathbf{D}^{\top}$, we have
\begin{equation}
\mathbf{D}^{\top}\mathbf{L} =\mathbf{D}^{\top}\mathbf{DK}\mathbf{I},
\label{eq: sr-formation-3}
\end{equation}
where $\mathbf{D}^{\top}\mathbf{D}$ is a selection matrix which is defined as
\begin{equation}
\label{eq: linear-operator}
\mathbf{D}^{\top}\mathbf{D}(\mathrm{x}, \mathrm{y}) = \left\{\begin{array}{ll} 1, & \mbox{}\ \mathrm{x} = \mathrm{y} = ds,\\0,
& \mbox{}\ $otherwise$, \end{array}\right.
\end{equation}
where $\mathrm{x}$ and $\mathrm{y}$ denote pixel locations; $d = \{1,\ldots, P\}$, and $P$ denotes the number of the pixels in $L$.
If $\mathbf{D}^{\top}\mathbf{D}(\mathrm{x}, \mathrm{x}) = 1$, we denote $\mathrm{x}$ as the un-decimated position.
The constraint~(\ref{eq: sr-formation-3}) indicates that the pixel value of $\mathrm{x}$ in ${L}$  is equal to the pixel value of $s\mathrm{x}$ in the blurred high resolution image
$\mathbf{B} = \mathbf{KI}$ at the un-decimated positions.
In the following, we will use the image formation constraint~(\ref{eq: sr-formation-3}) to guide our SR algorithm so that it can generate high-resolution images satisfying this constraint.
\begin{table}[!t]
\caption{\label{tab: network-parameters} Network parameters. ResBlock denotes the residual block~\cite{ResNet/CVPR16} which is used in~\cite{edsr}.
}
\centering
\begin{tabular}{lccccccc}
\toprule
Layers                  & Conv$_1$ & ResBlock & Conv$_2$ & Conv$_3$\\
\hline
Filter size             &  3         & 3     &3     &3    \\
Filter numbers          & 256        &256    &256   &3   \\
Stride                  &1           &1      &1     &1   \\
\bottomrule
\end{tabular}
\end{table}

\section{Proposed Algorithm}
The analysis above inspires us to use the image formation process to constrain the deep neural networks for SR.
Specifically, we first generate an intermediate HR image $I$ from a LR image $L$ by a deep neural network.
Then we apply the convolution kernel to $I$ and use pixel substitution
(Section~\ref{ssec: Pixel substitution}) to enforce the image formation constraint in the feedback step, as shown in Figure~\ref{fig: flow-chart}. 
In the following, we will explain the details of the proposed algorithm.

\subsection{Intermediate HR image estimation}
The effectiveness of using deep CNNs to super-resolve images has been extensively validated in SR problems.
Our goal here is not to propose a novel network structure but to develop a new framework to constrain the generated SR results using the image formation process.
Thus, we can use an existing network architecture, such as EDSR~\cite{edsr}, SRCNN~\cite{SRCNN/ECCV}, and VDSR~\cite{VDSR}.
In this paper, we use similar network architecture by~\cite{edsr} as our HR image estimation sub-network.
Figure~\ref{fig: flow-chart} shows the proposed network architecture for one stage of the proposed cascaded approach.
The parameters of the network are shown in Table~\ref{tab: network-parameters}, where we use $32$ ResBlocks and $0.1$ as the residual scaling factor~\cite{edsr}.

\subsection{Pixel substitution}
\label{ssec: Pixel substitution}
Let ${I}$ be the output of the HR image estimation sub-network. If ${I}$ is the ground truth HR image, the equality in the SR formation model~(\ref{eq: sr-formation-3}) strictly holds.
Thus, to enforce the intermediate HR image ${I}$ to be close to the ground truth HR image, we adopt the pixel substitution operation~\cite{shan/sr/sa08}.
Specifically, we first obtain the upsampled image $\mathbf{D}^{\top}\mathbf{L}$ by applying the upsampling matrix $\mathbf{D}^{\top}$ to the LR image $\mathbf{L}$.
Then we obtain the blurred intermediate HR image $\mathbf{B}$ and by applying the blur kernel $\mathbf{K}$ to the intermediate HR image $\mathbf{I}$.
The output of the pixel substitution operation is
\begin{equation}
\label{eq:pxiel:substitution}
\mathbf{\hat{B}}(\mathrm{x}) = \left\{\begin{array}{ll} \mathbf{D}^{\top}\mathbf{L}(\mathrm{x}), & \mbox{}\ \mathrm{x} = ds,\\ \mathbf{B} (\mathrm{x}),
& \ $otherwise$. \end{array}\right.
\end{equation}
Empirically, we find that the approximation scheme for image formation process converges well, as shown in the supplemental material.

\begin{table*}[!t]
  \caption{Quantitative evaluations of the state-of-the-art super-resolution methods on the benchmark datasets (Set5, Set14, B100, Urban100, Manga109, and DIV2K) in terms of PSNR and SSIM.
  }
\label{tab:psnr-sr}
\resizebox{.98\textwidth}{!}{
 \centering
 \begin{tabular}{lcccccccccc}
    \toprule
     &        &Set5            &Set14       &B100        &Urban100        &Manga109      &DIV2K (validation)\\
      Algorithms          & Scale                          &PSNR/SSIM       &PSNR/SSIM   &PSNR/SSIM   &PSNR/SSIM       &PSNR/SSIM     &PSNR/SSIM\\
    \hline
    Bicubic                                        &$\times 2$      &33.66/0.9299       &30.24/0.8688            &29.56/0.8431           &26.88/0.8403      &30.80/0.9339            &32.45/0.9040\\
    A+            &$\times 2$      &36.54/0.9544       &32.28/0.9056            &31.21/0.8863           &29.20/0.8938      &35.37/0.9680            &34.56/0.9330\\
    SRCNN         &$\times 2$      &36.66/0.9542       &32.45/0.9067            &31.36/0.8879           &29.50/0.8946      &35.60/0.9663            &34.59/0.9320\\
    FSRCNN        &$\times 2$      &37.05/0.9560       &32.66/0.9090            &31.53/0.8920           &29.88/0.9020      &36.67/0.9710            &34.74/0.9340\\
    VDSR          &$\times 2$      &37.53/0.9590       &33.05/0.9130            &31.90/0.8960           &30.77/0.9140      &37.22/0.9750            &35.43/0.9410\\
    LapSRN        &$\times 2$      &37.52 0.9591       &33.08/0.9130            &31.08/0.8950           &30.41/0.9101      &37.27/0.9740            &35.31/0.9442\\
    MemNet        &$\times 2$      &37.78/0.9597       &33.28/0.9142            &32.08/0.8978           &31.31/0.9195      &37.72/0.9740            &NA/NA\\
    DRCN          &$\times 2$      &37.63/0.9588       &33.04/0.9118            &31.85/0.8942           &30.75/0.9133      &37.57/0.9730            &35.45/0.940\\
    EDSR          &$\times 2$      &38.19/0.9609       &33.92/0.9195            &32.35/0.9019           &32.97/0.9358      &39.20/0.9783            &36.56/0.9485\\
    RDN           &$\times 2$      &38.24/0.9614       &\bf{34.01/0.9212}       &32.34/0.9017           &32.89/0.9353      &39.18/0.9780            &36.52/0.9483\\
    DBPN          &$\times 2$      &38.09/0.9600       &33.85/0.9190            &32.27/0.9000           &32.55/0.9324       &38.89/0.9775            &36.37/0.9475\\
    Ours          &$\times 2$      &\bf{38.26/0.9614}  &33.99/0.9200            &\bf{32.37/0.9020}      &\bf{33.09/0.9365}  &\bf{39.26/0.9784}       &\bf{36.60/0.9487}\\
    \hline
    Bicubic       &$\times 3$      &30.39/0.8682        &27.55/0.7742          &27.21/0.7385            &24.46/0.7349         &26.95/0.8556           &29.66/0.8310\\
    A+            &$\times 3$      &32.58/0.9088        &29.13/0.8188          &28.29/0.7835            &26.03/0.7973         &29.93/0.9120           &31.09/0.8650\\
    SRCNN         &$\times 3$      &32.75/0.9090        &29.30/0.8215          &28.41/0.7863            &26.24/0.7989         &30.48/0.9117           &31.11/0.8640\\
    FSRCNN        &$\times 2$      &33.18/0.9140        &29.37/0.8240          &28.53/0.7910            &26.43/0.8080         &31.10/0.9210           &31.25/0.8680\\
    VDSR          &$\times 3$      &33.67 0.9210        &29.78 0.8320          &28.83 0.7990            &27.14 0.8290         &32.01 0.9340           &31.76/0.8780\\
    LapSRN        &$\times 3$      &33.82/0.9227        &29.87/0.8320          &28.82/0.7980            &27.07/0.8280         &32.21/0.9350           &31.22/0.8600\\
    MemNet        &$\times 3$      &34.09/0.9248        &30.00/0.8350          &28.96/0.8001            &27.56/0.8376         &32.51/0.9369           &NA/NA\\
    DRCN          &$\times 3$      &33.82/0.9226        &29.76/0.8311          &28.80/0.7963            &27.15/0.8276         &30.97/0.8860           &31.79/0.8770\\
    EDSR          &$\times 3$      &34.68/0.9293        &30.52/0.8462          &29.26/0.8096            &28.81/0.8658         &{\bf{34.19}}/0.9485    &32.75/0.8933\\
    RDN           &$\times 3$      &34.71/0.9296        &30.57/0.8468          &29.26/0.8093            &28.80/0.8653         &34.13/0.9484           &32.73/0.8929\\
    Ours          &$\times 3$     &\bf{34.75/0.9298}    &\bf{30.61/0.8466}     &\bf{29.29/0.8102}        &\bf{28.97/0.8683}   &34.14/\bf{0.9490}      &\bf{32.79/0.8939}\\
    \hline
    Bicubic       &$\times 4$      &28.42/0.8104        &26.00/0.7027          &25.96 0.6675            &23.14/0.6577         &24.89/0.7866           &28.11/0.7750\\
    A+            &$\times 4$      &30.28/0.8603        &27.32/0.7491          &26.82/0.7087            &24.32/0.7183         &27.03/0.8510           &29.28/0.8090\\
    SRCNN         &$\times 4$      &30.48/0.8628        &27.50/0.7513          &26.90/0.7101            &24.52/0.7221         &27.58/0.8555           &29.33/0.8090\\
    FSRCNN        &$\times 4$      &30.72/0.8660        &27.61/0.7550          &26.98/0.7150            &24.62/0.7280         &27.90/0.8610           &29.36/0.8110\\
    VDSR          &$\times 4$      &31.35/0.8830        &28.02/0.7680          &27.29/0.0726            &25.18/0.7540         &28.83 0.8870           &29.82/0.8240\\
    LapSRN        &$\times 4$      &31.54/0.8850        &28.19/0.7720          &27.32/0.7270            &25.21/0.7560         &29.09/0.8900           &29.88/0.8250\\
    MemNe        &$\times 4$      &31.74/0.8893        &28.26/0.7723          &27.40/0.7281            &25.50/0.7630         &29.42/0.8942           &NA/NA\\
    DRCN          &$\times 4$      &31.53/0.8854        &28.02/0.7670          &27.23/0.7233            &25.14/0.7510         &28.97/0.8860           &29.83/0.8230\\
    EDSR          &$\times 4$      &32.48/0.8985        &28.80 0.7876          &27.72/0.7419            &26.65/0.8032         &31.03/0.9156           &30.73/0.8445\\
    RDN           &$\times 4$      &32.47/0.8990        &28.81/0.7871          &27.72/0.7419            &26.61/0.8028         &31.00/0.9151           &30.71/0.8442\\
    DBPN         &$\times 4$      &32.47/0.8980        &{\bf{28.82}}/0.7860   &27.72/0.7400            &26.38/0.7946         &30.91/0.9137           &30.66/0.8424\\
    Ours		  &$\times 4$     &\bf{32.56/0.8995}    &28.80/{\bf{0.7882}}   &\bf{27.73/0.7422}       &\bf{26.72/0.8053}    &\bf{31.06/0.9169}      &\bf{30.74/0.8450}\\
 \bottomrule
  \end{tabular}
  }
\end{table*}

\subsection{Cascaded training}
As the proposed algorithm consists of both intermediate HR image estimation and pixel substitution, we perform these two steps in a cascaded manner.
Let $\Theta_t$ denote the model parameters at stage (iteration) $t$, and $\{L^n, I_{gt}^n\}_{n = 1}^{N}$ denote a set of $N$ training samples.
We learn the stage-dependent model parameters $\Theta_t$ from~$\{L^n, I_{gt}^n\}_{n = 1}^{N}$ by minimizing the cost function
\begin{equation}
\label{eq: loss-function}
\mathcal{J}(\Theta_t) = \sum_{n = 1}^{N} \|{I}_t^{n}-I_{gt}^{n}\|_1,
\end{equation}
where $I_t^{n}$ is the output of the network at the $t$-th stage. Following~\cite{edsr}, we use $L_1$ norm as the loss function.
We minimize~(\ref{eq: loss-function}) to learn the model parameters $\Theta_t$ stage by stage from $t = 1, ..., T$.
%

\begin{figure*}[!t]
\centering
\begin{tabular}{c}
\includegraphics[width=0.96\linewidth, height = 0.26\linewidth]{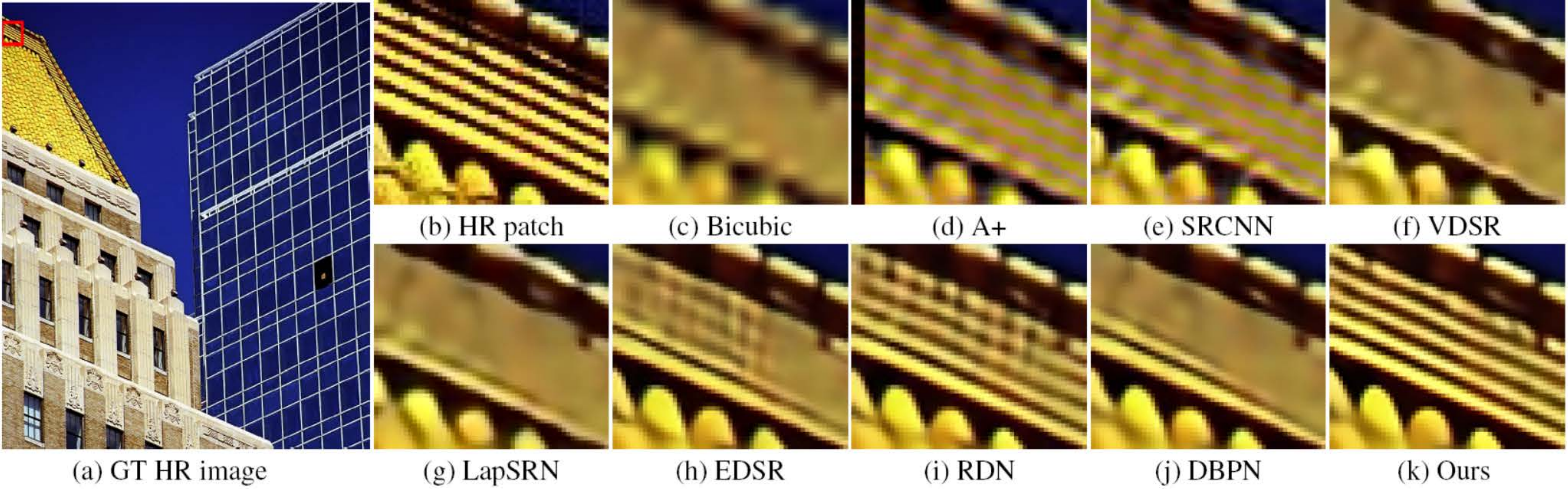} \\
\end{tabular}
\caption{Visual comparisons for $4\times$ SR from the Urban100 dataset. The proposed algorithm generates much better results with fine detailed structures.}%
\label{fig: sr4}
\end{figure*}

\begin{figure*}[!t]
\centering
\begin{tabular}{cccccccc}
\includegraphics[width=0.96\linewidth]{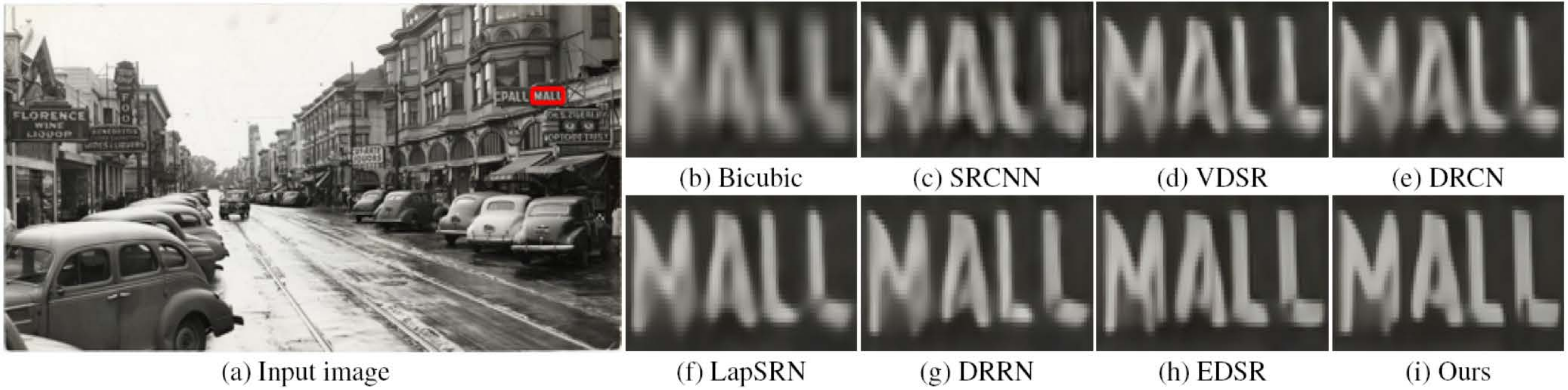} \\
\end{tabular}
\caption{Results on real example ($\times 4$). The proposed method recovers much clearer images with better detailed structures.}%
\label{fig: real-examples}
\end{figure*}
%

\section{Experimental Results}
\label{sec: experimental-results}
We examine the proposed algorithm using publicly available benchmark datasets and compare it to state-of-the-art single image SR methods.
Due to the extensive experiments performed, we only show a small portion of the results in the main paper. Please see the supplementary document for more results.
\subsection{Parameter settings and training data}
\label{ssec: Parameter settings and training data}
In the learning process, we use the ADAM optimizer~\cite{adam} with parameters $\beta_1 = 0.9$, $\beta_2 = 0.999$, and $\epsilon = 10^{-8}$.
The minibatch size is set to be $1$. The learning rate is initialized to be $10^{-4}$.
We use a Gaussian kernel in~(\ref{eq: sr-formation-3}) with the same settings used in~\cite{shan/sr/sa08}.
We empirically set $T = 3$ as a trade-off between accuracy and speed.
In the first stage, we use the same upsampling layer as~\cite{edsr} to upsample the features before the Conv$_3$ layer.

For fair comparisons, we first follow standard protocols adopted by existing methods (e.g.,~\cite{DBPN,edsr,RCAN,Zhang_2018_CVPR}) to generate LR images using bicubic downsampling from the DIV2K dataset~\cite{NTIRE2017} for training and use the Set5~\cite{Set5} as the validation test set.
Then, we evaluate the effectiveness of our algorithm when LR images are obtained with different image formation models of SR in Section~\ref{sec: analysis}.
%
We implement our algorithm based on the PyTorch version of~\cite{edsr}. The code and trained models are publicly available on the authors¡¯ websites.
\subsection{Comparisons with the state of the art}
To evaluate the performance of the proposed algorithm, we compare it against state-of-the-art algorithms
including A+~\cite{A+/accv15}, SRCNN~\cite{SRCNN/ECCV}, FSRCNN~\cite{Accelerating/dong}, VDSR~\cite{VDSR}, LapSRN~\cite{LapSR}, MemNet~\cite{MemNet}, DRCN~\cite{DRCN}, DRRN~\cite{DRRN}, EDSR~\cite{edsr}, RDN~\cite{Zhang_2018_CVPR}, and DBPN~\cite{DBPN}.
We use the benchmark datasets: Set5~\cite{Set5}, Set14~\cite{Set14}, B100~\cite{BSDS}, Urban100~\cite{SelfEx/sr}, Manga109~\cite{manga109}, and DIV2K (validation set)~\cite{NTIRE2017} to evaluate the performance.
These datasets contain different image diversities, e.g., the Set5, Set14, and B100 datasets consist of natural scenes;
Urban100 mainly contains urban scenes with details in different frequency bands; Manga109 is a dataset of Japanese manga; DIV2K (validation set) contains 100 natural images with 2K resolution.
%
We use the PSNR and SSIM to evaluate the quality of each recovered image.

Table~\ref{tab:psnr-sr} summarizes the quantitative results on these benchmark datasets for the upsampling factors of 2, 3, and 4.
Overall, the proposed method performs favorably against the state-of-the-art methods.
%


Figure~\ref{fig: sr4} shows SR results with a scale factor of 4 by the evaluated methods.
The results by the feed-forward models~\cite{SRCNN/ECCV,Accelerating/dong,VDSR,LapSR,edsr,Zhang_2018_CVPR} do not recover the structures well.
The DBPN algorithm~\cite{DBPN} adopts a feedback network to super-resolve images using information from the LR images.
However, this method needs to regenerate LR featrues from intermediate HR features.
Consequently, the information at un-decimated pixels would get lost, which makes it hard to estimate the details and structures.
The results in Figure~\ref{fig: sr4}(j) show that the structures of the images super-resolved by the DBPN method are not recovered well.
In contrast, the proposed method recovers finer image details and structures than the state-of-the-art algorithms.
{\flushleft \bf {Real examples.}}
We further evaluate our algorithm using real images (Figure~\ref{fig: real-examples}).
Our algorithm generates much clearer images with better detailed structures than those by the state-of-the-art methods~\cite{VDSR,SRCNN/ECCV,LapSR,edsr}.
For example, all the four letters in our result are legible, especially ``A'' (Figure~\ref{fig: real-examples}(i)).

\section{Analysis and Discussions}
\label{sec: analysis}
We have shown that enforcing the image formation constraint using pixel substitution leads to an algorithm that outperforms  state-of-the-art methods.
To better understand the proposed algorithm, we perform further analysis, compare it with related methods, and discuss its limitations.
%
\begin{figure}[!t]
\centering
\begin{tabular}{c}
\includegraphics[width=0.98\linewidth]{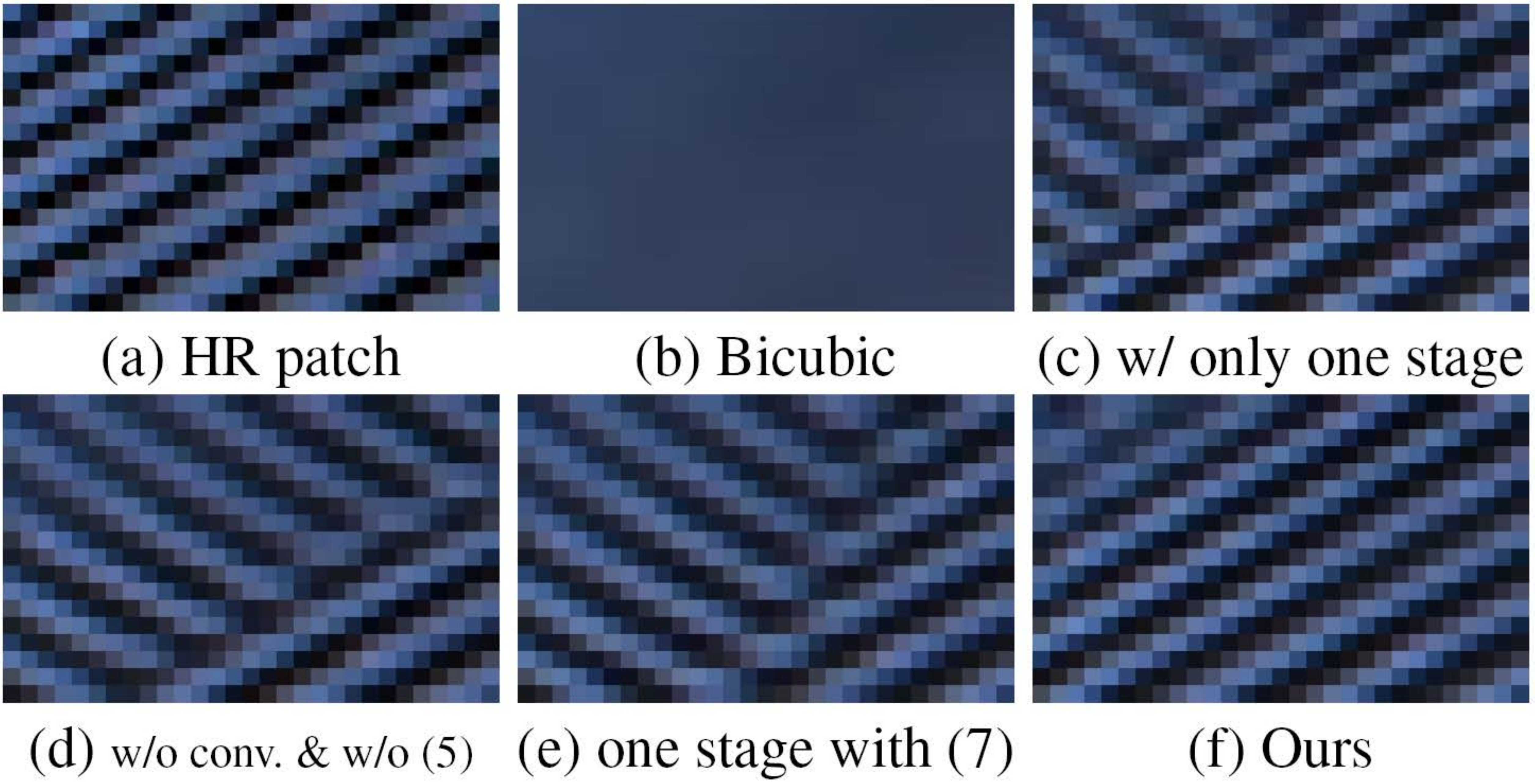}\\
\end{tabular}
\caption{Effectiveness of the image formation constraint ($\times$4). The image formation plays an important role for the details and structures estimation.}%
\label{fig: image-formation-effect}
\end{figure}

\begin{table}[!t]
  \caption{Effectiveness of the image formation constraint in SR with the scale factor 2. ``w/ only one stage" means the baseline method~\cite{edsr}.
  ``Stage 1", ``Stage 2", ``Stage 3" denote the intermediate results from stages 1, 2, and 3 of the proposed method, respectively. Other notations are detailed in the main context.
  }
   \label{tab: illustration-sr-psnr}
 \resizebox{0.47\textwidth}{!}{
 \centering
 \begin{tabular}{lcccccccccc}
    \toprule
   Dataset                                                 &Set5                                &B100                      &Urban100               &Manga109\\
                                                           &PSNR/SSIM                         &PSNR/SSIM                 &PSNR/SSIM              &PSNR/SSIM\\
    \hline
 w/o conv. \&~(\ref{eq:pxiel:substitution})                &38.20/0.9612                   &32.33/0.9017               &32.78/0.9347          &39.07/0.9780\\
 w/ only one stage                                         &38.19/0.9609                   &32.35/0.9019               &32.97/0.9358          &39.20/0.9783\\
 one stage with~(\ref{eq: loss-function-new})              &38.22/0.9613                   &32.33/0.9014               &32.82/0.9351          &39.00/0.9777\\
 Stage 1                                                   &38.23/0.9613                   &32.34/0.9017               &32.86/0.9350          &39.15/0.9782\\
 Stage 2                                                   &\bf{38.27/0.9614}              &32.36/{\bf{0.9020}}        &33.04/0.9362          &\bf{39.26/0.9784}\\
 Stage 3                                                   &38.26/\bf{0.9614}         &\bf{32.37/0.9020}          &\bf{33.09/0.9365}     &{\bf{39.26/0.9784}}\\
 \bottomrule
  \end{tabular}
  }
\end{table}

%
{\flushleft \bf{Effectiveness of the image formation constraint.}}
As our cascaded architecture uses a basic SR network several times, one may wonder whether the performance gains merely come from the use of a larger network.
To answer this question, we remove the pixel substitution step from our cascaded network architecture (denoted as ``w/o conv. \&~(\ref{eq:pxiel:substitution})") for fair comparisons.
The comparisons in Figure~\ref{fig: image-formation-effect}(d) and (f) demonstrate the benefit of using the image formation constraint in generating clearer images with finer details and structures.
We note that there is little performance improvement by simply cascading a basic SR network several times to increase the network capacity (Figure~\ref{fig: image-formation-effect}(d)).
The results in Table~\ref{tab: illustration-sr-psnr} show that using the image formation constraint of SR consistently improves SR results, which further demonstrates the effectiveness of this constraint.

\begin{figure}[!t]
\centering
\begin{tabular}{c}
\includegraphics[width=1\linewidth]{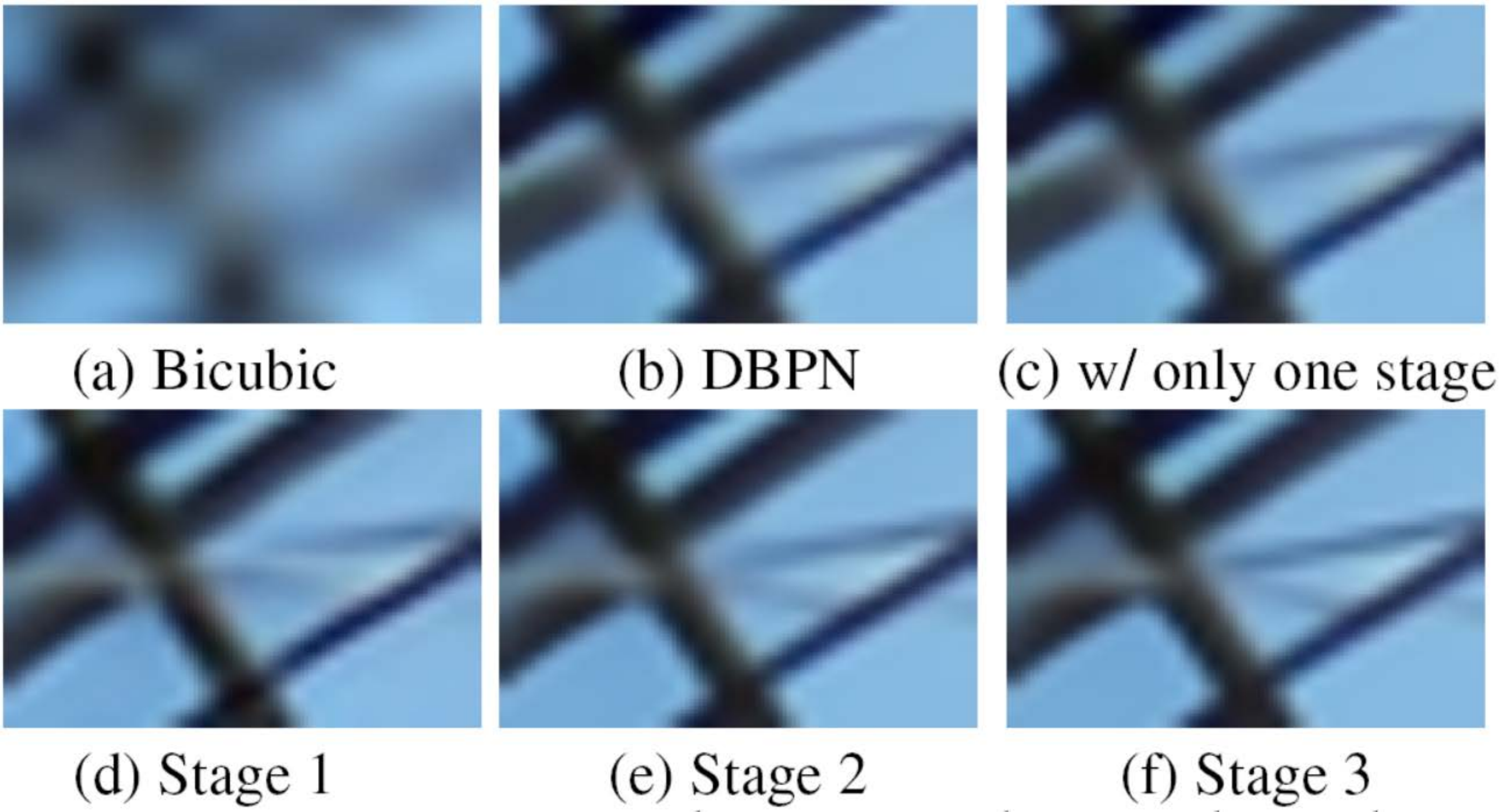} \\
\end{tabular}
\caption{Effectiveness of the proposed stage-dependent algorithm ($\times$4). (c) denotes the results with only one stage. (d)-(f) denote the intermediate HR images from stage 1, 2, and 3, respectively. }%
\label{fig: stage-optimization}
\end{figure}

Figure~\ref{fig: stage-optimization} shows some intermediate HR images from the proposed method.
We note that the structural details are better recovered with more stages. This further demonstrates
that using the image formation constraint in a deep CNN helps the restoration of the structural details.

As the proposed network architectures are similar to those used in~\cite{edsr}, the proposed algorithm with only one stage would reduce to the feed-forward model~\cite{edsr} to some extent.
Both the quantitative evaluations in Table~\ref{tab: illustration-sr-psnr} and comparisons in Figure~\ref{fig: image-formation-effect}(c) show that only using one feed-forward model does not generate high-quality HR images.

We further note that an alternative approach is to add the image formation model~(\ref{eq: sr-formation-convolution}) to the loss function to constrain the network training instead of using feedback scheme, where the new loss function is defined as
\begin{equation}
\label{eq: loss-function-new}
\ell_p(I; I_{gt}; L) = \|{I} - I_{gt}\|_1 + \lambda \|\downarrow^{s}(k\otimes I) - L\|_1,
\end{equation}
where $\lambda$ is a weight parameter. We empirically set $\lambda=0.01$ for fair comparisons in this paper. We denote this baseline method as ``one stage with~(\ref{eq: loss-function-new})".
We quantitatively evaluate this baseline model on the benchmark datasets.
Both the quantitative results in Table~\ref{tab: illustration-sr-psnr} and visual comparison (Figure~\ref{fig: image-formation-effect}(e)) demonstrate that adding the image formation loss to the overall loss function does not always improve the performance.

\begin{figure}[!t]
\centering
\begin{tabular}{c}
\includegraphics[width=0.96\linewidth]{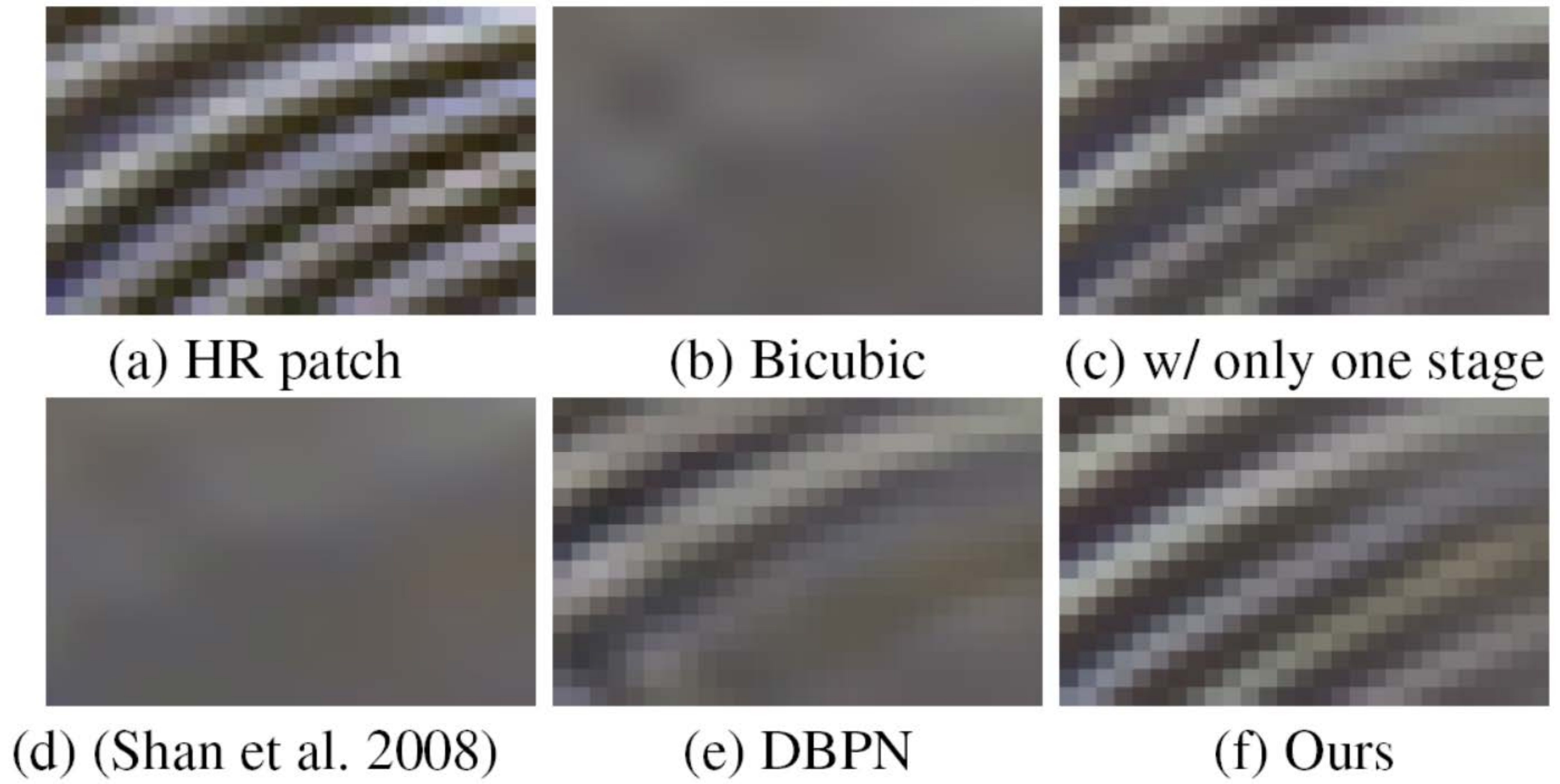} \\
\end{tabular}
\caption{Comparisons of the results by different back-projection methods ($\times$4). The DBPN method based on the iterative back-projection algorithm~\cite{IBP} is less effective for the edges restoration
due to the additional downsampling operation.}%
\label{fig: ibp-algorithm}
\end{figure}

%
{\flushleft \bf{Closely-related methods.}}
Several notable methods~\cite{DBPN,shan/sr/sa08} improve the back-projection algorithm~\cite{IBP} for single image SR.
The DBPN algorithm~\cite{DBPN} extends the back-projection method~\cite{IBP} using a deep neural network.
It needs a downsampling operation after obtaining intermediate HR images in the feedback stage.
As the information at un-decimated pixels of the intermediate HR images may be lost due to the downsampling operation, DBPN is less effective at recovering details and structures (Figure~\ref{fig: ibp-algorithm}(e)).
The method~\cite{shan/sr/sa08} first proposes pixel substitution to enforce the image formation constraint in an iterative optimization scheme.
However, this method cannot effectively restore the edges and textures (Figure~\ref{fig: ibp-algorithm}(d)) because only the sparsity of gradient prior is used.
In contrast, our algorithm uses pixel substitution to constrain the deep CNN.
%
Both the edges and textures are well recovered (see Figure~\ref{fig: ibp-algorithm}(f)).

We further examine whether the estimated HR images satisfy the image formation constraint. To this end, we apply the image formation to the estimated HR images to generate the LR images
and use the PSNR and mean squared error (MSE) as the metrics.
The MSE values in Table~\ref{tab: illustration-sr-psnr-lr} indicate that the results generated by the proposed method satisfy the image formation model well.
\begin{table}[!t]
  \caption{Evaluations of the regenerated LR images ($\times 2$).
  }
   \label{tab: illustration-sr-psnr-lr}
 \resizebox{0.47\textwidth}{!}{
 \centering
 \begin{tabular}{lcccccccccc}
    \toprule
   Dataset                                                 &Set5                 &B100                  &Urban100                  &Manga109    \\
                                                           &PSNR/MSE             &PSNR/MSE              &PSNR/MSE                  &PSNR/MSE    \\
    \hline
 DBPN                                          &61.60/0.0462         &60.07/0.0704          &59.00/0.0944               &60.41/0.0658      \\
 Ours                                                      &\bf{72.06/0.0045}    &\bf{66.29/0.0264}     &\bf{65.04/0.0360}          &\bf{67.56/0.0189}   \\

 \bottomrule
  \end{tabular}
}
\end{table}

%
{\flushleft \bf{Robustness to general degradation models of SR.}}
We have shown that using the image formation constraint can make the deep CNNs more compact thus facilitating the SR problem when the degradation model is approximated by the
Bicubic downsmapling operation in Section~\ref{ssec: Parameter settings and training data}.
%
%
We further evaluate the proposed algorithm when the degradation model is approximated by the Bicubic downsmapling with noise. To generate LR images for training,
we add Gaussian noise to each LR image used in Section~\ref{ssec: Parameter settings and training data}, where the noise level ranges from 0 to 10\%.
Table~\ref{tab:psnr-noise} shows that our algorithm is robust to image noise due to the cascaded optimization method. More results about other degradation models are included in the supplementary document.

All above results on both synthetic and real-world images demonstrate that the proposed algorithm can generalize well even though the image formation constraint is based on known blur kernels.
%

\begin{figure}[!t]
\centering
\begin{tabular}{c@{}c@{}c}
\includegraphics[width=0.3\linewidth]{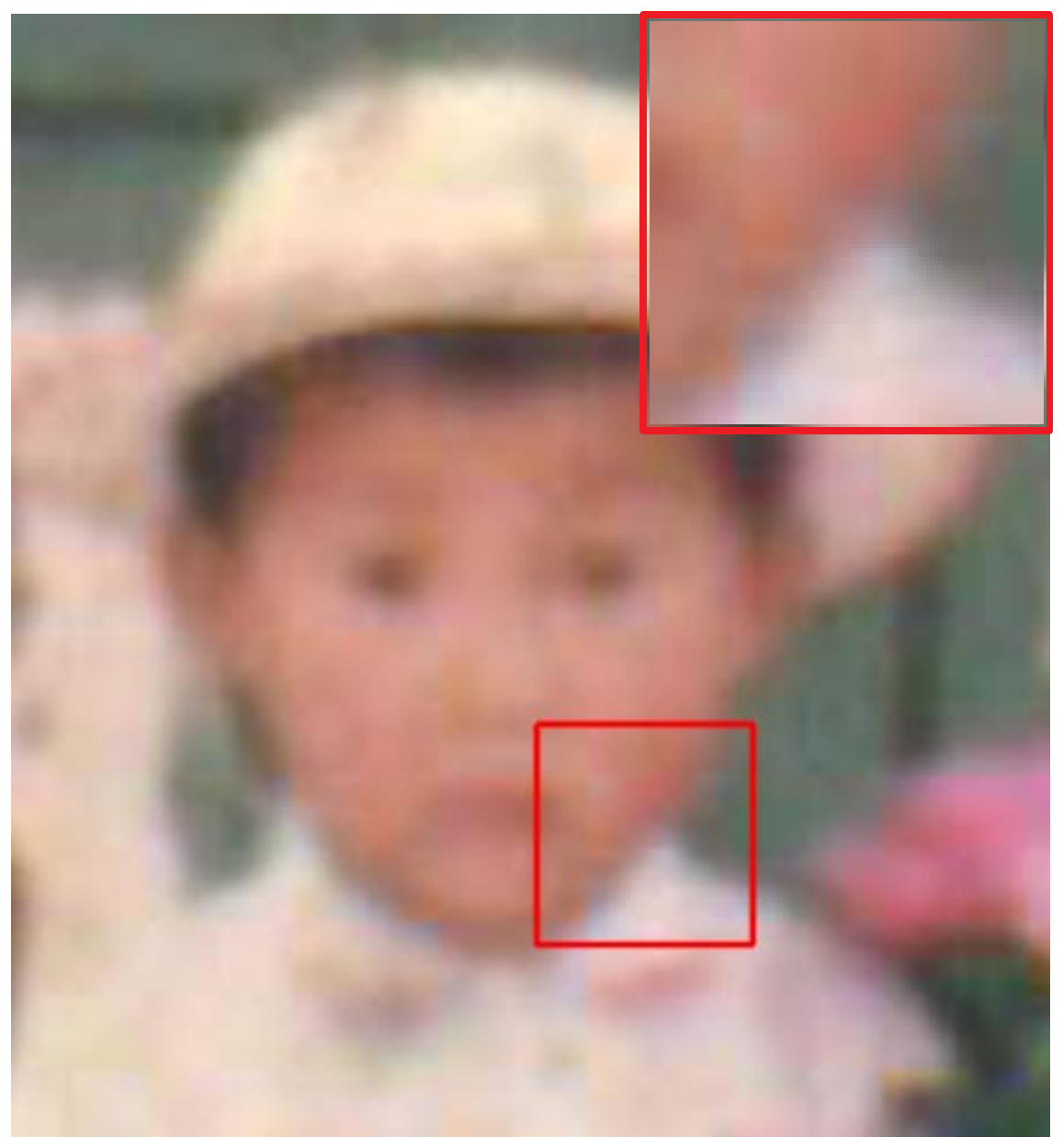} &
\includegraphics[width=0.3\linewidth]{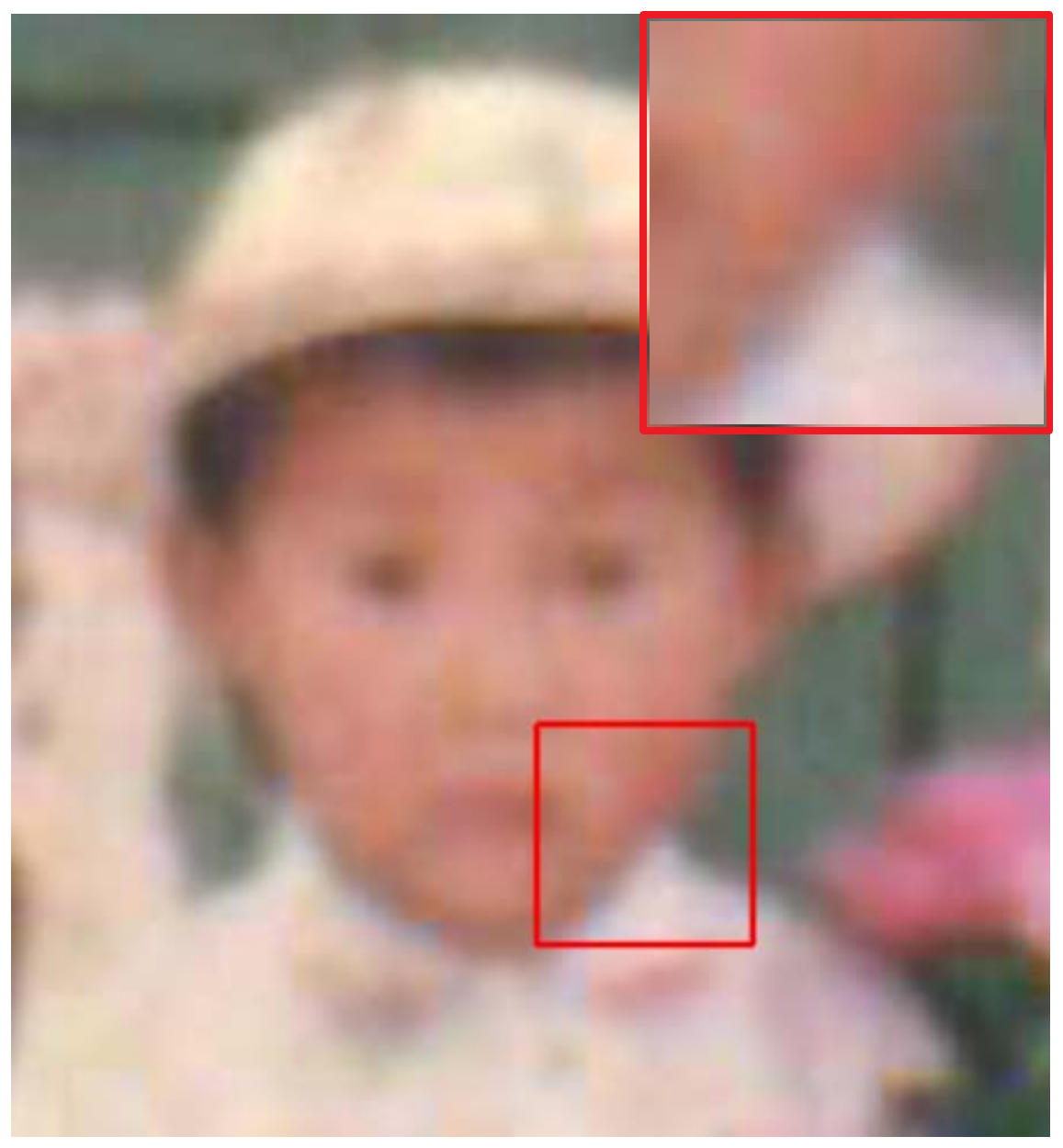} &
\includegraphics[width=0.3\linewidth]{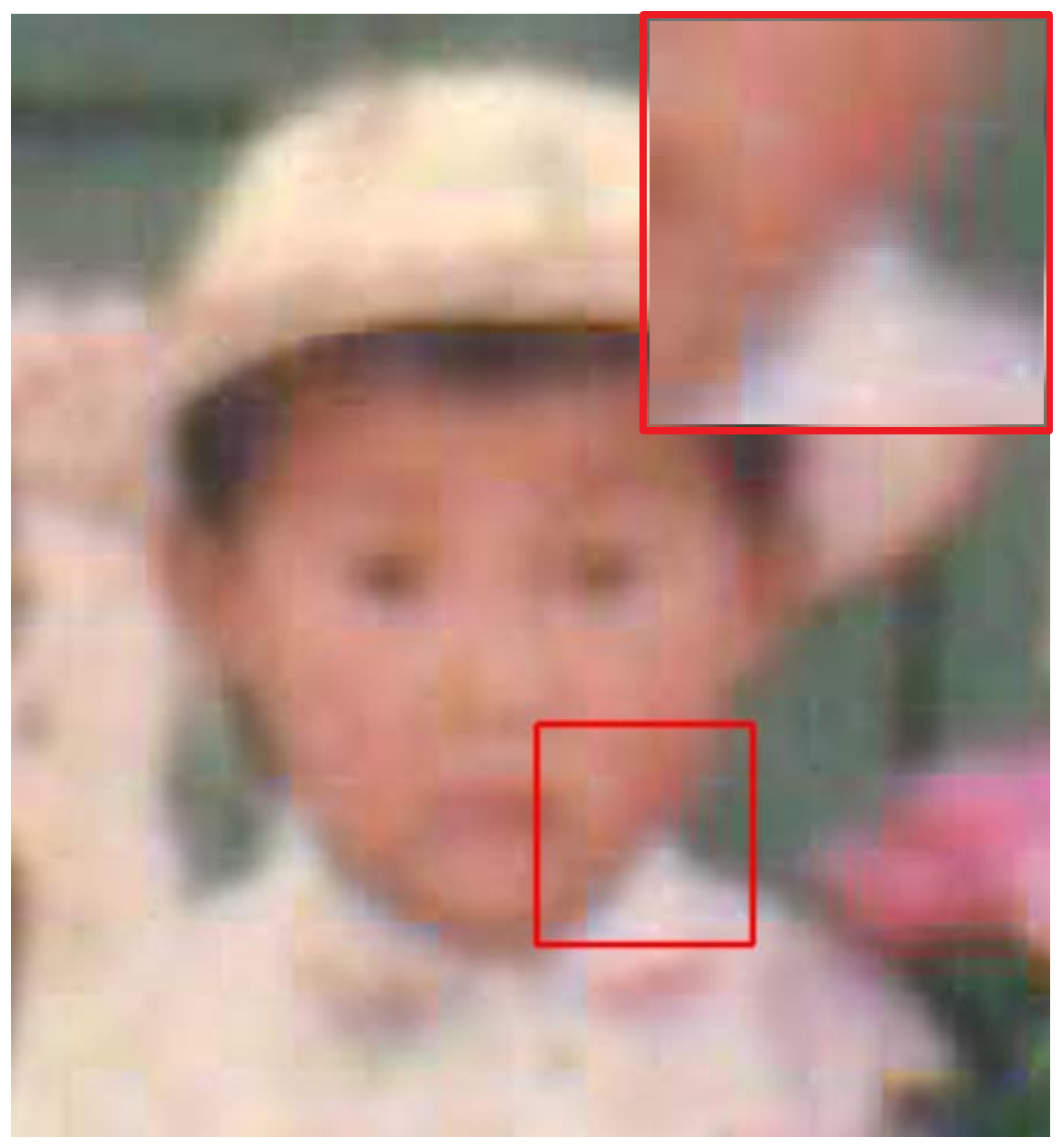} \\
(a) Input image &(b) VDSR & (c) Ours\\
\end{tabular}
\caption{The proposed algorithm is less effective when the image formation of SR does not hold. Using the image formation of SR to super-resolve images with JPEG compression artifacts would exaggerate the artifacts (Best viewed on high-resolution display with zoom-in).
}
\label{fig: limitation}
\end{figure}

\begin{table}[!t]
  \caption{Results ($\times 2$) on ``Set5" with noisy input images.
  }
   \label{tab:psnr-noise}
 \resizebox{0.46\textwidth}{!}{
 \centering
 \begin{tabular}{lcccccccccccc}
    \toprule
    Noise level            &1\%    & 2\%         &3\%      &4\%\\
    \hline
     EDSR                           &38.19/0.9610            &35.96/0.9387                &35.03/0.9272                &34.30/0.9179            \\
     RDN                 &38.16/0.9609            &35.69/0.9382                &35.02/0.9272                &34.20/0.9176         \\
     Ours                                       &\bf{38.21/0.9611}       &\bf{36.01/0.9389}           &\bf{35.06/0.9275}           &\bf{34.34/0.9184}       \\
 \bottomrule
  \end{tabular}
}
\end{table}
%
%
{\flushleft \bf{Running time performance.}}
As our algorithm uses a cascaded architecture, it increases the computation.
We examine the running time of the proposed algorithm and compare it with state-of-the-art methods on the Set5 dataset, as shown in Table~\ref{tab: run-time}.
The proposed algorithm  takes slightly more running time compared with the feed-forward models, e.g.,~\cite{VDSR,edsr}.
The proposed algorithm is about $3$ times faster than the feedback DBPN method~\cite{DBPN}.
{\flushleft \bf{Limitations.}}
As our algorithm uses the known image formation of SR to approximate the unknown degradation model of SR,
it is less effective when this approximation does not hold.
%
%
Figure~\ref{fig: limitation} shows an example with significant JPEG compression artifacts, where the image formation model of SR does not approximate the degradation caused by the image compression well.
Our algorithm exacerbates the compression artifacts, while the results by the feed-forward models have few artifacts.
Building the compression process into the network architecture is likely to reduce these artifacts.

\begin{table}[!t]
  \caption{Running time performance on SR with a scale factor of $2$.
  }
   \label{tab: run-time}
\resizebox{0.48\textwidth}{!}{
 \centering
 \begin{tabular}{cccccc}
    \toprule
    Methods                  &VDSR           &EDSR           &RDN           &DBPN         &Ours \\
    \hline
 Avg. running time (/s)         &0.88              &1.16          &2.01            & 6.81         & 2.21 \\
 \bottomrule
  \end{tabular}
  }
\end{table}

\section{Concluding Remarks}

We have introduced a simple and effective super-resolution algorithm that exploits the image formation constraint.
The proposed algorithm first uses a deep CNN to estimate an intermediate HR image and
then uses pixel substitution to enforce the intermediate HR image satisfy the image formation model at the un-decimated pixel positions.
Our cascaded architecture can be applied to existing feed-forward super-resolution networks.
Both quantitative and qualitative results show that the proposed algorithm performs favorably against state-of-the-art methods.

{\flushleft \bf{Acknowledgement.}} This work has been supported in part by the National Natural Science
Foundation of China (Nos. 61922043, 61872421, 61732007) and the Natural Science Foundation of Jiangsu Province (No. BK20180471).
\bibliography{egbib}
\bibliographystyle{aaai}

\end{document}